\documentclass[fleqn,usenatbib]{mnras}

\usepackage{newtxtext,newtxmath}

\usepackage[T1]{fontenc}

\DeclareRobustCommand{\VAN}[3]{#2}
\let\VANthebibliography\thebibliography
\def\thebibliography{\DeclareRobustCommand{\VAN}[3]{##3}\VANthebibliography}

\usepackage{graphicx}	
\usepackage{amsmath}	
\usepackage{makecell}
\usepackage{xcolor}
\usepackage{xspace}

\newcommand{\revision}[1]{{\color{black}#1}}
\newcommand{\newrevision}[1]{{\color{black}#1}}

\usepackage{threeparttable}
\usepackage{longtable}
\newcommand{\fref}[1]{Fig.~\ref{#1}}
\newcommand{\tref}[1]{Table~\ref{#1}}

\newcommand{\cref}[1]{Chapter~\ref{#1}}
\newcommand{\sref}[1]{Section~\ref{#1}}
\newcommand{\aref}[1]{Appendix~\ref{#1}}
\usepackage{tabularx} 
\usepackage{hyperref}

\usepackage{amsmath} 

\newcolumntype{L}{>{\raggedright\arraybackslash}X}

\usepackage{orcidlink}

\usepackage{pdflscape}

\title[SN Ia environments with DustPedia -- I.]{The UV-to-FIR environments of nearby Type Ia supernovae with the DustPedia Galaxy Catalogue -- I. Local versus global host properties}

\author[L. Kelsey]{
L.~Kelsey$^{1}$\thanks{E-mail: lisa.kelsey@ast.cam.ac.uk}\thanks{Leverhulme Trust and Isaac Newton Trust Early Career Fellow} \orcidlink{0000-0003-0313-0487}\\
$^{1}$ Institute of Astronomy and Kavli Institute for Cosmology, University of Cambridge, Madingley Road, Cambridge CB3 0HA, UK   \\
}

\date{Accepted XXX. Received YYY; in original form ZZZ}

\pubyear{\the\year{}}

\begin{document}
\label{firstpage}
\pagerange{\pageref{firstpage}--\pageref{lastpage}}
\maketitle

\begin{abstract}
The ongoing use of Type Ia Supernovae (SNe Ia) as standardisable candles relies on \revision{an additional correction} for the relationship between their standardised brightnesses and host galaxy properties. \revision{Currently, t}his correction is built on host stellar mass, typically derived from optical-only SED fitting, which recent analyses suggest may not fully capture the underlying environmental properties. This relationship is stronger when considering the regions closest to the SNe, rather than the galaxies as a whole. I use a sample of SNe Ia in DustPedia galaxies and \texttt{CIGALE} SED fitting to compare the local environments of \revision{90} nearby SNe Ia against the global properties of their hosts. The UV-to-FIR coverage of DustPedia enables improved dust constraints, and helps break the age-dust degeneracy that limits optical-only studies; to my knowledge this coverage has not been previously applied to the sub-galactic regions around SNe Ia. Stellar mass and dust mass are lower locally simply because the aperture encloses less of the galaxy; however, for colour, sSFR, and dust attenuation, local values differ from the host average by more than measurement uncertainty alone allows, with attenuation showing the strongest discrepancy \revision{($p=6.3\times10^{-5}$)}. Only mass-weighted age is consistent locally and globally. Galaxy-integrated properties therefore do not capture the true conditions at the SN location. SN siblings (sharing a host galaxy) only emphasise this effect. Siblings sampling the same galaxy structure, such as two SNe in spiral arms, have consistent environments, whilst others diverge. Standardising on global properties may therefore mask the differences that drive the brightness-environment relationship. 
\end{abstract}




\section{Introduction}

Type Ia Supernovae (SNe Ia), the explosive deaths of carbon-oxygen white dwarfs in binary systems, have proved vital in aiding our understanding of the universe, most notably through the discovery of its accelerating expansion \citep{Perlmutter1999,Riess1998}. As standardisable candles with a naturally low absolute magnitude dispersion ($\sim0.35$\,mag) which can be tightened further to $\sim0.14$\,mag \citep{Scolnic2018} using the ``brighter-slower'' \citep{Rust1974,Pskovskii1977,Phillips1993} and ``brighter-bluer'' \citep{Riess1996,Tripp1998} relations, SNe Ia have secured their place as a cornerstone of modern cosmology. 

Despite this, there is plenty we are yet to fully understand about SNe Ia. Of these, perhaps the most crucial to solve is the relationship between standardised brightness and SN Ia host galax\revision{y properties}, which is now the largest systematic uncertainty in SN cosmology \citep{Vincenzi2024,Popovic2026}. SNe Ia are found in all types of galaxy \citep[e.g.][]{Sullivan2003}, making them common and easy to observe across a wide redshift range. Whilst this is a benefit for building large samples, it also means that the SN Ia population is drawn from the full diversity of stellar environments, and therefore any environmental dependence of standardised brightness feeds directly into cosmology.

The most well-studied correlation is referred to as the ``mass step'': SNe Ia in more massive hosts are brighter after standardisation than those in less massive hosts by $\sim0.06$\,mag when split at $\log(M_*/\mathrm{M_\odot}) = 10$ \citep{Kelly2010,Lampeitl2010,Sullivan2010}. This relationship has been observed in other galaxy parameters, such as specific star formation rate \citep{Sullivan2010,Rigault2013,Rigault2020}, colour \citep{Roman2018,Kelsey2021,Kelsey2023}, metallicity \citep{DAndrea2011,Childress2013}, stellar population age \citep{Rose2019} and morphology \citep{Pruzhinskaya2020}. The consensus is not that any one environmental property is the cause of the dispersion; instead, these properties act as proxies or tracers for the true driver \revision{(see \citealt{Briday2022}, who compare the standardisation efficiency of different tracers)}. The underlying cause has been variously attributed to progenitor astrophysics, to dust, or to a combination of the two \citep[e.g.][]{Childress2013,Rigault2013,Rigault2020,Kim2018,BS2021,Thorp2021,Thorp2022,Briday2022,Wiseman2022,Wiseman2023,Kelsey2023,Popovic2023,Popovic2024,Grayling2024,Grayling2025, Hayes2025,Ginolin2026,Magee2026,Sarin2026}. 

Making a correction based on one tracer removes much of the step, but residual brightness offsets remain when subsequently considering other properties \citep{Kelsey2023}, and recent analyses find that the standardisation parameters themselves, particularly the colour-luminosity coefficient $\beta$, vary significantly with host galaxy position in the SFR--$M_*$ plane \citep{Ramaiya2025}. Despite this growing body of evidence, SN cosmology still applies only a single global mass-based correction. This reduces the complex nature of galaxies to essentially one number, yet galaxies at a fixed stellar mass span a range in star formation rate, colour and morphology \citep[e.g.][]{Baldry2006,Schawinski2014,Tacchella2019}. Logically therefore, a single global property cannot encode the environment each SN actually experienced.

Global photometry measurements are also dominated by the most luminous regions of a galaxy, which may not resemble the SN site. One option is therefore to instead measure the properties of the galaxy environment more locally around each SN site. Both spectroscopic studies, through $\textrm{H}\alpha$ and integral field spectroscopy, and photometric studies of regions around SN sites, have shown that these sub-galactic, ``local'' environments provide a clearer view of the SN progenitor environment and often show stronger relationships with SN Ia light-curve parameters and brightnesses than their global counterparts \citep{Rigault2013, Rigault2015, Rigault2020, MorenoRaya2016, MorenoRaya2016b, Jones2018, Roman2018, Kim2018, Galbany2018, Kim2019, Rose2019, Kelsey2021, Kelsey2023}.

However, global stellar mass persists as the default correction, in part because it is a property that can be inferred relatively cleanly from the optical broadband photometry that SN surveys already obtain as part of their routine observations, which are designed around optimal SN discovery. Some host studies rely solely on this survey imaging \citep[e.g. DES;][]{Smith2020,Kelsey2021}, whilst others supplement it with additional photometry \revision{including in the near-infrared} \citep[NIR; e.g.][]{Sullivan2010,Roman2018,Uddin2020,Ponder2021}, though rarely beyond. \revision{Recent work has begun to push host SED fitting further into the infrared, from the mid-infrared (\citealt{Murakami2026}, Tweddle et al., in prep) to the far-infrared \citep{Ramaiya2025}, better constraining dust-obscured star formation and the separation of passive from dusty star-forming hosts.} Host spectroscopy is hard to obtain and expensive at the scale of the SN samples required for cosmology, and deriving stellar population properties from optical-only photometry can be unreliable due to the inability to accurately constrain age, metallicity, dust and star formation \citep[e.g.,][]{Worthey1994,Papovich2001,Gallazzi2005,Conroy2013}. Local properties are more difficult still. At the high redshifts covered by typical cosmology surveys, local aperture photometry can become comparable to the size of an individual galaxy, even when considering relatively high-resolution deep-stacked data from ground-based optical surveys \citep[e.g. as was the case for DES,][]{Kelsey2021,Kelsey2023}. As a result, where local and global measurements have been compared directly, it has been with the same restricted data, for the same few properties.

Whilst this analysis considers very nearby SNe Ia, particularly when compared to the wide redshift ranges of modern cosmological surveys, these events are valuable for understanding the interplay between SNe Ia and their most immediate environments. Their hosts can be resolved in far greater detail than those at higher redshift, with wavelength coverage extending into the FIR that constrains dust and star formation directly. This enables measurements of how local environment departs from the global host with a precision higher-redshift samples cannot reach. 

In this analysis I focus on the physical properties of the local environments of Type Ia supernovae within DustPedia host galaxies. I compare these local properties to the global host galaxy values and examine the case of sibling supernovae, where multiple events in the same host provide a controlled test of environmental variation. I do not attempt to correlate these properties with SN Ia light-curve parameters such as stretch or colour, as many of the events in the sample are historic and lack well-calibrated light curves. Such correlations, for the subset of SNe Ia with modern photometric coverage, are deferred to a future publication in this series. In \sref{data} I introduce the sample and data available for this study, before outlining the photometric methods, selection cuts and SED fitting technique in \sref{methods}. I present the results for the full sample in \sref{environments} and for the subset of siblings (multiple SNe Ia in the same galaxy) in \sref{siblings}. I discuss these results in \sref{Discussion}, before summarising the key findings and conclusions in \sref{Summary}.
\section{Data}\label{data}

\subsection{DustPedia Galaxy Sample}\label{dustpedia}

The DustPedia project \citep{Davies2017} is built around the legacy of the \textit{Herschel Space Observatory}, with the aim of understanding dust in the nearby universe. The catalogue consists of 875 galaxies, comprising the majority of extended ($D_{25} > 1$\arcmin) galaxies within $v_\mathrm{helio} < 3000$\,km\,s$^{-1}$ ($z \lesssim 0.01$) that were observed by \textit{Herschel}. For each galaxy, \citet{Clark2018} constructed a homogeneous multi-wavelength database from \textit{GALEX} (FUV, NUV), SDSS ($ugriz$), 2MASS ($JHK_s$), \textit{WISE} (3.4, 4.6, 12, 22\,$\mu$m), \textit{Spitzer} IRAC (3.6, 4.5, 5.8, 8.0\,$\mu$m) and MIPS (24, 70, 160\,$\mu$m), \textit{Herschel} PACS (70, 100, 160\,$\mu$m) and SPIRE (250, 350, 500\,$\mu$m), and \textit{Planck}, achieving up to 42 bands spanning over five orders of magnitude in wavelength for these galaxies, with an average galaxy possessing photometry in 25 bands. All images were reduced in a standardised way, with foreground stars masked, and aperture-matched photometry performed consistently across all bands with compatible uncertainty estimates for each galaxy as outlined in \citet{Clark2018}.

This combination of extended, nearby galaxies and wide wavelength coverage means that the DustPedia catalogue provides an excellent sample to investigate the differences between global host galaxy and local environments of SNe Ia. The angular resolution of even the coarsest band used in this analysis (\textit{Herschel} SPIRE 350\,$\mu$m, FWHM $= 24.9$\arcsec) corresponds to physical scales of a few kpc at these distances, sufficient to isolate the SN site from the rest of the host galaxy. This is not possible for higher-redshift SN samples where any local measurement is effectively a global one. 

The wavelength coverage provided by the DustPedia catalogue is vital for such an analysis. A galaxy spectral energy distribution (SED) is comprised of components from different physical processes which appear at different wavelengths \citep[see e.g.][for a review]{Conroy2013}. UV and optical bands constrain the unattenuated stellar populations and the degree of dust attenuation, while the FIR bands directly trace the dust emission. Without FIR coverage, SFR estimates in particular can be biased\revision{, a concern \citet{Kim2024} raise for local measurements from optical photometry}; \citet{Ramaiya2025} show that extending SN\,Ia host galaxy photometry from the optical into the MIR and FIR yields significantly different SED-derived star formation rates compared to optical-only analyses, due to degeneracies between dust attenuation and stellar age that can only be broken with longer-wavelength data (see also \citealt{Ramaiya2026} for a radio-based alternative for hosts lacking FIR coverage). 

\subsection{SN Ia Sample Selection}

To identify SNe Ia hosted by galaxies in the DustPedia catalogue, I compiled a comprehensive list of transients from the Transient Name Server (TNS\footnote{\url{https://www.wis-tns.org/}}) and the Open SN Catalogue \citep[OSC;][]{OSC2017} and cross-matched against the 875 DustPedia galaxies. 

The TNS catalogue was obtained in two parts: a static archive covering events recorded prior to the TNS era (containing 6568 transients discovered before early 2015) and a bulk download of TNS public objects, last updated on 21st May 2026.\footnote{\url{https://www.wis-tns.org/content/tns-getting-started}.} Please note that additional transients reported to the TNS or classified as SN Ia between then and the submission of this manuscript are not included in this sample. Galaxy coordinates and names were drawn from the DustPedia sample catalogue. TNS SNe were associated with DustPedia galaxies via two methods. First, the pre-TNS catalogue was matched with the galaxy sample on galaxy name, yielding 244 candidate associations of all SN types. This method was applied to the pre-TNS data only, as host names are not provided in the TNS bulk download files. Second, a positional cone search was performed on all remaining unmatched transients (both pre- and post-TNS catalogues) with a purposefully large matching radius of 10~arcmin, adopted to account for SNe projected at large angular distances from their host. For example, SN\,2011fe in NGC\,5457 (M\,101) lies 4.6~arcmin from the galaxy centre. Transients were then filtered to only those with ``Ia'' in their classifications, retaining all subtypes and peculiar classes (e.g.\ SN Ia-pec, SN Ia-91bg-like, SN Ia-91T-like, SN Iax[02cx-like]).

To ensure completeness, particularly for older or less well-catalogued events, an independent cross-match against the OSC was performed. The OSC was filtered to transients with a claimed type containing ``Ia''. Host galaxy names were matched against DustPedia galaxy names after normalising for formatting and naming convention differences (e.g., whitespace between prefix and catalogue number, ``Messier'' versus ``M'' designations). Where multiple host galaxies were listed for a single transient, each was checked independently. Most transients were already present in the TNS host-name matched sample, but 10 were unique to the OSC consisting predominantly of older events (pre-1990) without a reported host galaxy on the TNS.

The TNS and OSC samples were then merged. Where a transient appeared in both catalogues, the TNS coordinates were adopted. Host associations were manually reviewed by eye and by cross-checking with the literature. For example, SN\,2005X and SN\,2024bhx are removed from the sample as they are associated with a satellite galaxy near NGC\,4353 which is not in the DustPedia catalogue, instead of NGC\,4353 itself.

Following manual review, a sample of 107 confirmed SNe Ia hosted by 92 distinct DustPedia galaxies remains, \revision{as outlined in \tref{tab:selection}}. Of these host galaxies, 11 are associated with two or more SNe Ia \citep[SN siblings, \citealt{Brown2014}; see e.g.][]{Kelsey2024}.

\begin{table}
\centering
\caption{Sample selection based on DustPedia host photometry flags \revision{and local aperture size requirement.}}
\label{tab:selection}
\begin{tabular}{lcc}
\hline
Selection & Galaxies & SNe~Ia \\
\hline
Confirmed SN~Ia in DustPedia hosts & 92 & 107 \\
Contamination-flagged photometry & $-9$ & $-12$ \\
Insufficient SED coverage & $-3$ & $-3$ \\
\hline
After photometry coverage cuts & 80 & 92 \\
$3$\,kpc local aperture radius $>$ host galaxy $R_{25}$ & $-2$ & $-2$ \\
\hline
Final sample & 78 & 90 \\
\hline
\end{tabular}
\end{table}

\section{Methods}\label{methods}

\subsection{Global Photometry}

In order to compare the local, sub-galactic parameters for each SN location to the global parameters derived for the whole galaxy consistently, I use the galaxy photometry reported by DustPedia \citep{Clark2018}. The DSS, \textit{Planck}, \textit{Herschel} SPIRE $500\,\mu\textrm{m}$ and \textit{Spitzer} MIPS $160\,\mu\textrm{m}$ are removed as outlined in \sref{processing} to match the resolution requirements for the local aperture photometry.

Each filter of DustPedia galaxy aperture-matched photometry carries a data quality flag, indicating contamination (c/C), artefacts (a/A) and insufficient sky coverage (n/N), with lower- and upper-case denoting minor and major issues respectively \citep{Clark2018}. Following \citet{Nersesian2019}, I exclude from the global fit any band carrying a major flag, retaining bands flagged as minor. The major flags and adopted approach to them are as follows: 
\begin{itemize}
    \item Contamination flags indicate that the global integrated photometry is a blend of the target and a nearby source, such as a nearby galaxy or poorly-subtracted foreground star, compromising the global properties obtained from a fit to such data. These cases typically impact so many bands for a galaxy that all bands receive the ``C'' flag. Whilst affected galaxies (9) and their associated SNe (12; this flag removed NGC\,1316, a host of four SNe Ia) were removed from the sample to allow for a fair global vs local environment comparison, I note that this flag illustrates a key benefit of local over global environments. Were a SN located in a galaxy merger environment, the stellar population properties at the SN site may be dramatically different from the host average, resulting in a more meaningful interpretation of its environment. 
    \item Artefact flags (satellite trails, saturation, etc.) can be present in individual bands when they fall within the photometric aperture. For bands affected by major artefact flags (A), the affected band was dropped from the global SED fit for that galaxy. Applying this after the above contamination cut, five hosts had major artefacts, each in one to three \textit{Spitzer} or 2MASS bands. On visual inspection, some of these artefacts clearly fell near to the SN site, so for consistency, these bands were also removed from the local photometry rather than treating each local aperture individually. 
    \item Null-coverage flags on the DustPedia global photometry indicate that the observation did not provide full coverage of the source area in the band, resulting in poor measurement of the flux and background \citep[see][for a full discussion of the percentage coverage resulting in a major/minor flag]{Clark2018}. This impacts \textit{Spitzer} and \textit{Herschel} PACS most strongly due to their small maps. Bands impacted by a major null-coverage flag (N) were removed from the global fits. This does not impact the local photometry, as the photometry extraction code automatically excludes uncovered pixels. 
\end{itemize}
\revision{To detail explicitly, the cut using contamination flags removes galaxies from the sample, whilst the artefact and null-coverage cuts remove individual bands from individual galaxies. The number of bands entering the SED fit therefore varies from host to host,} \newrevision{as is standard for FUV-FIR SED fitting analyses \citep[e.g.][]{Nersesian2019}.} After these cuts, it is additionally required that the surviving photometry bands have sufficient coverage across the SED, particularly $0.35 \leq \lambda/\mu\textrm{m} \leq 3.6$ and the $60 \leq \lambda/\mu\textrm{m} \leq 500$ region covering the peak of the dust emission, following \citet{Nersesian2019}, \newrevision{ensuring that the key SED features are retained to break the dust-age degeneracy, and thus trust the environmental parameters derived.} This \newrevision{requirement} removes three further galaxies from the sample: ESO\,428-014, NGC\,3972 and NGC\,7213.  

After these cuts, as summarised in \tref{tab:selection}, the sample comprises 80 host galaxies with 92 SNe Ia. 

\subsection{Local Photometry}

\subsubsection{Image Processing}\label{processing}

To process the DustPedia imaging data to obtain local aperture photometry for each SN in the sample, all available imaging data from the DustPedia archive \citep{Clark2018} for the hosts were retrieved, spanning UV to FIR wavelengths (see details in \sref{dustpedia}). Following \citet{Clark2018}, DSS imaging was excluded due to its non-linear photographic density units. \textit{Planck}, \textit{Herschel} SPIRE $500\,\mu\textrm{m}$ and \textit{Spitzer} MIPS $160\,\mu\textrm{m}$ were additionally removed as they all have too coarse pixel scales and angular resolution for spatially-resolved analyses.\footnote{For the spatial details for each band of DustPedia data, please refer to Table~1 from \citet{Clark2018}.}

Background subtraction \citep[required for all DustPedia imaging data, see][section 2.2]{Clark2018} and variance map generation were performed using a custom \texttt{Python} pipeline based on methods from \texttt{piXedfit} \citep{Abdurrouf2021}. For all images, background subtraction was performed using a 2D background model constructed using the \texttt{photutils Background2D} function \citep{Bradley2025}, with a $3\sigma$ clipping threshold, median background estimator, and a grid box size of one-tenth of the image dimensions in each axis following the method in \texttt{piXedfit}. The default pixel exclusion threshold of 10 per cent per box was sufficient for the majority of the images, but for a subset of \textit{Spitzer} images, where the data coverage is particularly sparse, the default threshold was insufficient and was increased incrementally in steps of 10 per cent until background estimation succeeded, with most requiring a threshold of no more than 30 per cent. For these cases, per-pixel \textit{Spitzer} uncertainty maps were available, such that the variance maps were constructed directly from these maps and are unaffected by the modified background estimation.

Variance maps were constructed for each image. Although \texttt{piXedfit} provides instrument-specific functions for constructing variance maps from raw survey data (e.g. \texttt{var\_img\_GALEX}, \texttt{var\_img\_2MASS}), these require instrument-specific header keywords that were not preserved in the DustPedia \texttt{.fits} data products. Instead, where uncertainty maps were available from the DustPedia archive, these were converted to variance maps. For all remaining bands, the RMS background noise map produced by the background subtraction process was used as the uncertainty estimate. In both cases, variance maps were produced using the \texttt{piXedfit var\_img\_from\_unc\_img} function.

Following background subtraction and variance map generation, point spread function (PSF) matching and spatial reprojection of the multiband imaging data was performed using a modified version of the \texttt{piXedfit} \texttt{images\_processing} class. The standard \texttt{piXedfit} implementation assumes instrument-native pixel units for each survey (e.g., counts per second for \textit{GALEX}, nanomaggies for SDSS, etc.) and applies instrument-specific unit conversions automatically. This class was modified to account for the fact that all DustPedia imaging data are provided in standardised units of Jy\,pixel$^{-1}$.

PSF matching was performed by convolving each image to the resolution of the band with the largest PSF for each galaxy, using convolution kernels from the \texttt{piXedfit} Google Drive repository\footnote{\url{https://drive.google.com/drive/folders/1pTRASNKLuckkY8_sl8WYeZ62COvcBtGn}} based on \citet{Aniano2011}. 

\subsubsection{Local Aperture Photometry}

To measure the local photometry in each SN region, a common physical aperture size must be defined. Following \citet{Roman2018} and \citet{Kelsey2021}, the maximum FWHM was converted to the smallest useful aperture radius of $1\sigma$ seeing using $\textrm{FWHM} = 2\sqrt{2\textrm{ln}2} \approx 2.355\sigma$ assuming a Gaussian PSF. The majority of the sample (77 of 80 galaxies) were convolved to the \textit{Herschel} SPIRE $350\,\mu\textrm{m}$ beam (FWHM $= 24.9$\arcsec, $\sigma_\textrm{PSF} = 10.6$\arcsec), while the remaining three galaxies (NGC\,4454, NGC\,5426, NGC\,5427) lack \textit{Herschel} SPIRE 350 coverage and were instead convolved to the \textit{WISE} W4 beam (FWHM $= 11.9$\arcsec, $\sigma_\textrm{PSF} = 5.0$\arcsec). 

\begin{figure}
    \centering
    \includegraphics[width=\columnwidth]{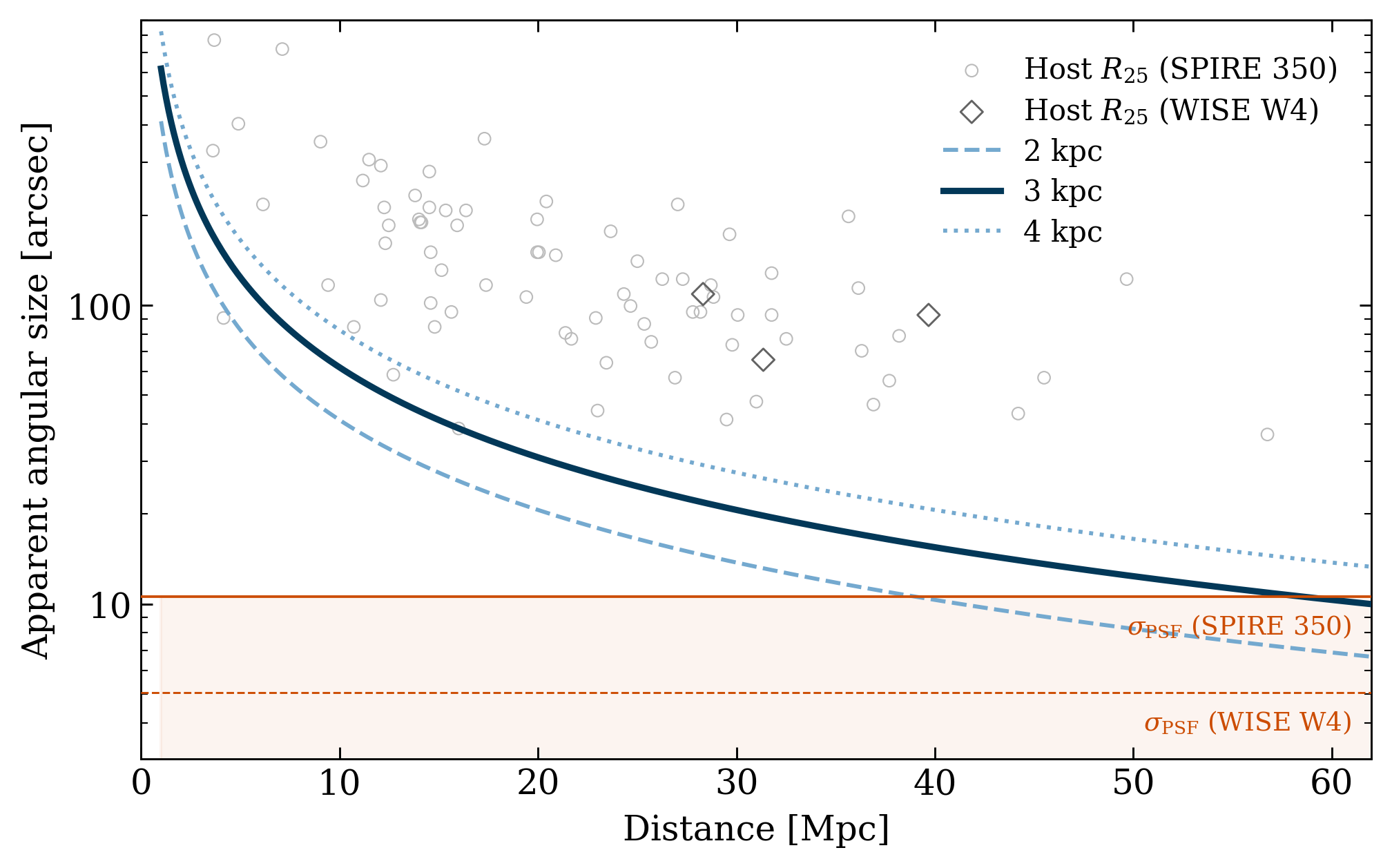}
    \caption{Apparent angular size of circular aperture radii of 2, 3, and 4\,kpc as a function of redshift-independent distance from the DustPedia catalogue. Open circles show host galaxy $R_{25}$ values for galaxies convolved to the \textit{Herschel} SPIRE 350\,$\mu$m beam, while open diamonds mark the subset convolved to \textit{WISE} W4 (galaxies lacking \textit{Herschel} SPIRE 350 coverage). The solid and dashed orange lines and shaded region mark $\sigma_{\mathrm{PSF}}$ for \textit{Herschel} SPIRE 350 (10.6\,arcsec) and \textit{WISE} W4 (5.0\,arcsec) respectively. A 3\,kpc aperture radius (solid dark curve) was adopted, which remains above $\sigma_{\mathrm{PSF}}$ for all 80 host galaxies \revision{passing the photometry coverage cuts}.}
    \label{fig:ap_vs_dist}
\end{figure}

\begin{figure*}
    \centering
    \includegraphics[]{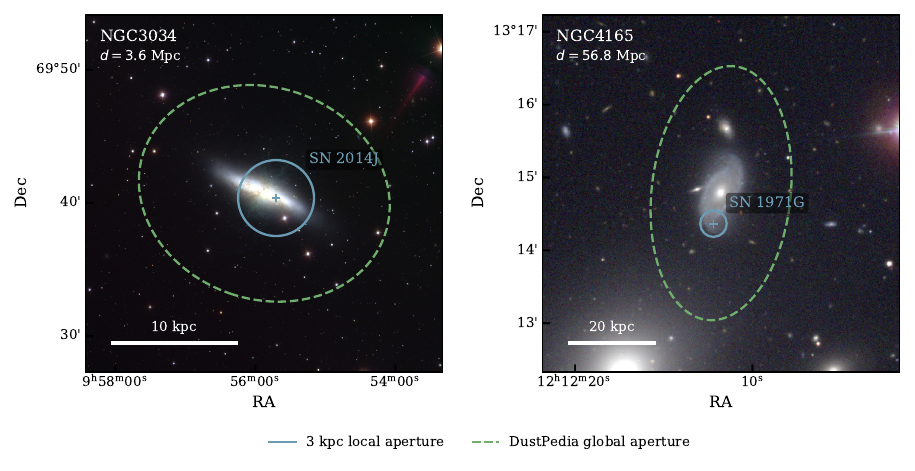}
    \caption{SDSS $gri$ false-colour composite images of two host galaxies in the sample, illustrating the range of distances and aperture sizes. \textit{Left:} NGC\,3034 (M\,82), the nearest host in the sample, hosting SN\,2014J. \textit{Right:} NGC\,4165, the most distant host, hosting SN\,1971G. Blue circles indicate the 3\,kpc physical apertures used for local photometry at each SN\,Ia site, with SN positions marked by crosshairs. \revision{The dashed green ellipse shows the DustPedia global aperture.} The white bar in each panel shows the physical scale.}
    \label{fig:eg_apertures}
\end{figure*}

In \fref{fig:ap_vs_dist}, the apparent size of different physical apertures as a function of distance is presented. A 3\,kpc radius remains above $\sigma_\textrm{PSF}$ for all 80 host galaxies, and below the isophotal radius $R_{25}$ (from HyperLeda via the DustPedia catalogue; \citealt{Makarov2014}) for all but two, meaning it probes regions smaller than the galaxy size but larger than the PSF. \revision{For the two exceptions, NGC\,4415 (where the aperture falls marginally at $R_{25}$ within 0.3 per cent), and NGC\,4424 (where it exceeds $R_{25}$), the 3\,kpc aperture is comparable to the size of each galaxy itself, so a `local' measurement cannot be meaningfully distinguished from a global one. I thus remove these two hosts and their associated SNe from the analysis, leaving \textbf{90 SNe Ia} across \textbf{78 host galaxies}, as outlined in \tref{tab:selection}.} \revision{A fixed local aperture of 3\,kpc is adopted for all SNe, so that local measurements are directly comparable across the sample. This is consistent with the fixed physical scales of $\sim$1--4\,kpc used across the SN Ia local environment literature \citep[e.g.][]{Rigault2013, Jones2018, Roman2018, Kelsey2021, Kelsey2023}.} Examples of this aperture for the nearest and furthest galaxy in the sample are presented in \fref{fig:eg_apertures}. I briefly investigate a 1\,kpc aperture radius for a subset of the data in \aref{aptest}.

Photometry is extracted in all bands simultaneously using the \texttt{photometry\_within\_aperture} function from \texttt{piXedfit}'s \texttt{piXedfit\_images} module, which sums the flux densities of all pixels enclosed within the circular aperture and propagates uncertainties in quadrature from the variance maps. The resulting flux densities, natively in $F_\lambda$ (erg\,s$^{-1}$\,cm$^{-2}$\,\AA$^{-1}$), are converted to $F_\nu$ in mJy using piXedfit's \texttt{convert\_flux\_unit} function for compatibility with the chosen SED fitting method (see \sref{SEDfitting}). \revision{The resulting local photometry for all 90 SNe is provided as online supplementary material.}

\subsection{SED Fitting}\label{SEDfitting}

SEDs of the SN Ia host galaxies and local environments are modelled using Code Investigating GALaxy Emission \citep[\texttt{CIGALE};][]{Noll2009, Boquien2019}.\footnote{\href{https://cigale.lam.fr/2025/10/06/version-2025-1/}{CIGALE Version 2025.1}} \texttt{CIGALE} constructs model SEDs by combining modules for the star-formation history, stellar populations, nebular emission, dust attenuation, and dust emission, assuming that the energy absorbed by dust at short wavelengths equals that re-emitted in the IR, providing self-consistent estimates of stellar mass, star formation rate, and dust mass. 

\texttt{CIGALE} uses a Bayesian analysis to derive probability distribution functions (PDFs) for the physical properties of each environment, with the likelihood-weighted mean and standard deviation of the PDF adopted as the estimate and its uncertainty \citep{Boquien2019}. \revision{For each model in the grid, the $\chi^2$ between the model and the observed photometry is computed, and the model is assigned a likelihood $\exp(-\chi^2/2)$. These likelihoods weight the contribution of every model to the PDF of each property, so the observed flux uncertainties propagate into the width of each PDF, as do degeneracies between models that fit the photometry equally well. Following the \texttt{CIGALE} default, an additional 10 per cent relative uncertainty is added in quadrature to all observed photometric errors to account for unknown systematics in the photometry and models \citep{Noll2009}.} 

The same models are adopted as \citet{Nersesian2019} (which in turn is based on \citealt{Hunt2019}), who fitted the global SEDs of the full DustPedia sample, to ensure consistency between the physical properties derived for the SN Ia host sample and the published DustPedia catalogue values. I note again here that I do not use all the filters available in DustPedia due to the requirement on pixel scale and angular resolution for a local analysis, so there may be small differences between these global properties and those presented by DustPedia. \revision{For the 78 hosts in common, my global estimates and the published DustPedia values show median offsets consistent with zero, with a scatter below 0.06~dex in stellar mass, star formation rate and specific star formation rate, and 0.03~mag in dust attenuation, rising to 0.18~dex in dust mass.} The model parameters are presented in full in \citet{Nersesian2019}, but for clarity for the reader I present here a brief summary of this setup with associated parameter grid in \tref{tab:cigale_params}.

\begin{table*}
\centering
\caption{Parameter grid used for computing the  \texttt{CIGALE}  models following \citet{Nersesian2019}.}
\label{tab:cigale_params}
\begin{threeparttable}
\begin{tabular}{llp{8cm}}
\hline\hline
Parameter & & Value \\
\hline
\multicolumn{3}{l}{\textit{Star-formation history} \hfill \texttt{sfhdelayedbq}} \\
& e-folding time, $\tau_\mathrm{main}$ (Myr) & 500, 750, 1100, 1700, 2600, 3900, 5800, 8800, 13\,000, 20\,000 \\
& Galaxy age, $t_\mathrm{gal}$ (Myr) & 2000, 4500, 7000, 9500, 12\,000 \\
& Burst/quench age, $t_\mathrm{flex}$ (Myr) & 200 \\
& SFR ratio, $r_\mathrm{SFR}$ & 0.01, 0.0316, 0.1, 0.316, 1.0, 3.16, 10.0 \\
\hline
\multicolumn{3}{l}{\textit{Stellar population model} \hfill \texttt{bc03}} \\
& IMF & Salpeter \\
& Stellar metallicity, $Z$ & 0.02 \\
\hline
\multicolumn{3}{l}{\textit{Nebular emission} \hfill \texttt{nebular}} \\
& Ionisation parameter, $\log U$ & $-2.0$ \\
& Gas-phase metallicity, $Z_\mathrm{gas}$ & 0.02 \\
& Electron density, $n_e$ (cm$^{-3}$) & 100 \\
& Ly continuum escape fraction, $f_\mathrm{esc}$ & 0.0 \\
& Ly continuum dust absorption fraction, $f_\mathrm{dust}$ & 0.0 \\
\hline
\multicolumn{3}{l}{\textit{Dust attenuation} \hfill \texttt{dustatt\_calzleit}} \\
& Colour excess of young population, $E(B{-}V)_\mathrm{young}$ (mag) & 0.0, 0.005, 0.0075, 0.011, 0.017, 0.026, 0.038, 0.058, 0.087, 0.13, 0.20, 0.29, 0.44, 0.66, 1.0 \\
& Old-to-young colour excess ratio, $E(B{-}V)_\mathrm{old} / E(B{-}V)_\mathrm{young}$ & 0.25, 0.50, 0.75 \\
& Attenuation curve slope modifier, $\delta$ & $-0.5$, $-0.25$, 0.0 \\
& UV bump amplitude & 0.0 \\
\hline
\multicolumn{3}{l}{\textit{Dust emission model} \hfill \texttt{themis}} \\
& Small hydrocarbon mass fraction, $q_\mathrm{hac}$ & 0.02, 0.06, 0.10, 0.14, 0.17, 0.20, 0.24, 0.28, 0.32, 0.36, 0.40 \\
& Minimum radiation field intensity, $U_\mathrm{min}$ & 0.1, 0.15, 0.3, 0.5, 0.8, 1.2, 2.0, 3.5, 6, 10, 17, 30, 50, 80 \\
& Radiation field power-law slope, $\alpha$ & 2.0 \\
& Dust fraction heated in PDR, $\gamma$ & 0.0, 0.001, 0.002, 0.004, 0.008, 0.016, 0.031, 0.063, 0.13, 0.25, 0.5 \\
\hline
\end{tabular}
\begin{tablenotes}
   \item This results in a total of 80\,041\,500 models. 
\end{tablenotes}
\end{threeparttable}
\end{table*}

A flexible delayed SFH is adopted which allows for bursts of star formation through the \texttt{sfhdelayedbq} module. The SFR as a function of time is given by:
\begin{equation}
    \mathrm{SFR}(t) \propto
    \begin{cases}
        t \times \exp(-t/\tau_\mathrm{main}), & t \leq t_\mathrm{flex} \\
        r_\mathrm{SFR} \times \mathrm{SFR}(t = t_\mathrm{flex}), & t > t_\mathrm{flex},
    \end{cases}
    \label{eq:sfh}
\end{equation}
where $\tau_\mathrm{main}$ is the e-folding time of the main stellar population, and $t_\mathrm{flex}$ marks the onset of a burst or quench episode with amplitude given by the ratio $r_\mathrm{SFR}$: $r_\mathrm{SFR} < 1$ corresponds to quenching, $r_\mathrm{SFR} > 1$ to a burst, and $r_\mathrm{SFR} = 1$ to an unmodified delayed SFH.

The stellar emission is computed from the \citet{BC2003} models assuming a \citet{Salpeter1955} initial mass function (IMF) and solar metallicity ($Z = 0.02$). Nebular emission components are modelled using the updated \textsc{cloudy} \citep{Ferland1998,Ferland2013} templates from \citet{VV2021}, with a constant ionisation parameter ($U$).

The stellar and nebular emission is attenuated using a power-law-modified starburst curve \citep{Calzetti2000}, extended to shorter wavelengths by \citet{Leitherer2002}:
\begin{equation}
    A(\lambda) = \left[ A(\lambda)_\mathrm{SB} \times \left(\frac{\lambda}{550~\mathrm{nm}}\right)^\delta \right] \times \frac{E(B{-}V)_{\delta=0}}{E(B{-}V)_\delta},
\end{equation}
where $\delta$ modifies the slope of the attenuation curve. A UV bump is not included ($D_\lambda = 0$) following \citet{Nersesian2019} as this would require additional UV data to be constrained that is not available for this sample.

Dust emission is modelled using \texttt{THEMIS} \citep[The Heterogeneous Evolution Model for Interstellar Solids;][]{Jones2013,Jones2017, Kohler2015}. \texttt{THEMIS} is built from laboratory-measured optical properties of amorphous hydrocarbon and amorphous silicate materials and parametrises the dust emission using three quantities: the mass fraction of small hydrocarbon solids $q_\mathrm{hac}$ (analogous to $q_\mathrm{PAH}$ in the \citealt{DraineandLi2007} model, with $q_\mathrm{PAH} \approx q_\mathrm{hac}/2.2$), the minimum radiation field intensity $U_\mathrm{min}$, and the fraction of dust heated in photodissociation regions $\gamma$.

All sources are assigned a redshift of zero for SED fitting, using instead the luminosity distance provided via redshift-independent distances from the DustPedia archive \citep{Davies2017} compiled from the HyperLeda database \citep{Makarov2014}, which overrides the distances computed from redshift in \texttt{CIGALE}, avoiding the inaccuracies of redshift-derived distances for nearby galaxies whose peculiar velocities dominate over the Hubble flow, which could impact the scaling of derived parameters.

\begin{figure*}
    \centering
    \includegraphics[width=\textwidth]{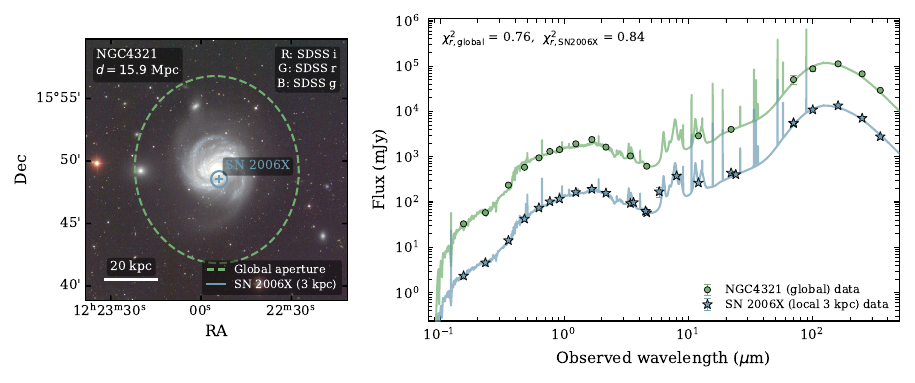}
    \caption{\textit{Left:} SDSS $gri$ false-colour composite of NGC\,4321 (M\,100), the host galaxy of SN\,2006X. The dashed green ellipse shows the DustPedia global photometric aperture, while the solid blue circle marks the 3\,kpc local aperture centred on the SN\,Ia site. \textit{Right:} Best-fit \texttt{CIGALE} SEDs for the global (green) and local (blue) apertures, with observed photometry shown as data points and the model spectra as solid lines.}
    \label{fig:best_SED}
\end{figure*}

The total parameter grid produces 80\,041\,500 models, identical to the grid used by \citet{Nersesian2019} for the full DustPedia sample. The SED fitting was carried out in \texttt{pdf\_analysis} mode on the Cambridge Service for Data Driven Discovery (CSD3),\footnote{\url{https://www.hpc.cam.ac.uk}} parallelised over 16 CPU cores. An example \texttt{CIGALE} best-fitting SED for both a global and local aperture is presented in \fref{fig:best_SED}. \revision{The resulting local and global property estimates for the full sample are provided as online supplementary material.}

\section{Environments of SNe Ia}\label{environments}

\texttt{CIGALE} fitting provides a wide range of parameters to explore for the SN host galaxies. In this analysis, I primarily focus on stellar mass ($M_*$), $u-r$ colour, specific star formation rate ($\mathrm{sSFR}$), mass-weighted age ($\textrm{Age}_{MW}$), dust attenuation ($A_V$), and dust mass ($M_\mathrm{dust}$). Together, these span the properties that feature prominently in the SN Ia host galaxy literature. $M_*$ is the basis of the widely used mass step correction \citep{Kelly2010,Sullivan2010,Lampeitl2010}; $u-r$ colour \citep{Roman2018,Kelsey2021,Kelsey2023} and $\mathrm{sSFR}$ \citep{Rigault2013,Rigault2020} trace the recent star-formation activity; $\textrm{Age}_{MW}$ probes the stellar population age \citep{Rose2019,Wiseman2023}, which has featured heavily in recent discussions about progenitor-age effects \citep{Wiseman2026,Murakami2026}; and the dust properties, $A_V$ and $M_\mathrm{dust}$, probe the host dust content central to the debate on whether the host-dependence is driven by dust or intrinsic colour \citep{BS2021,GG2021,Meldorf2023,Duarte2023,Popovic2024}. \revision{Metallicity is not considered, with both the stellar and gas-phase metallicities ($Z$ and $Z_\mathrm{gas}$) fixed at solar in the model grid (\tref{tab:cigale_params}) following \citet{Nersesian2019}. Broadband photometry alone cannot reliably constrain metallicity, which is best measured spectroscopically \citep[see][for a review]{Conroy2013}.} $\mathrm{sSFR}$ is derived from the \texttt{CIGALE} SFR and $M_*$, and $u-r$ from the model fluxes with propagated uncertainties; the remaining properties are taken directly from the Bayesian estimates. 

\begin{figure*}
    \centering
    \includegraphics[width=\linewidth]{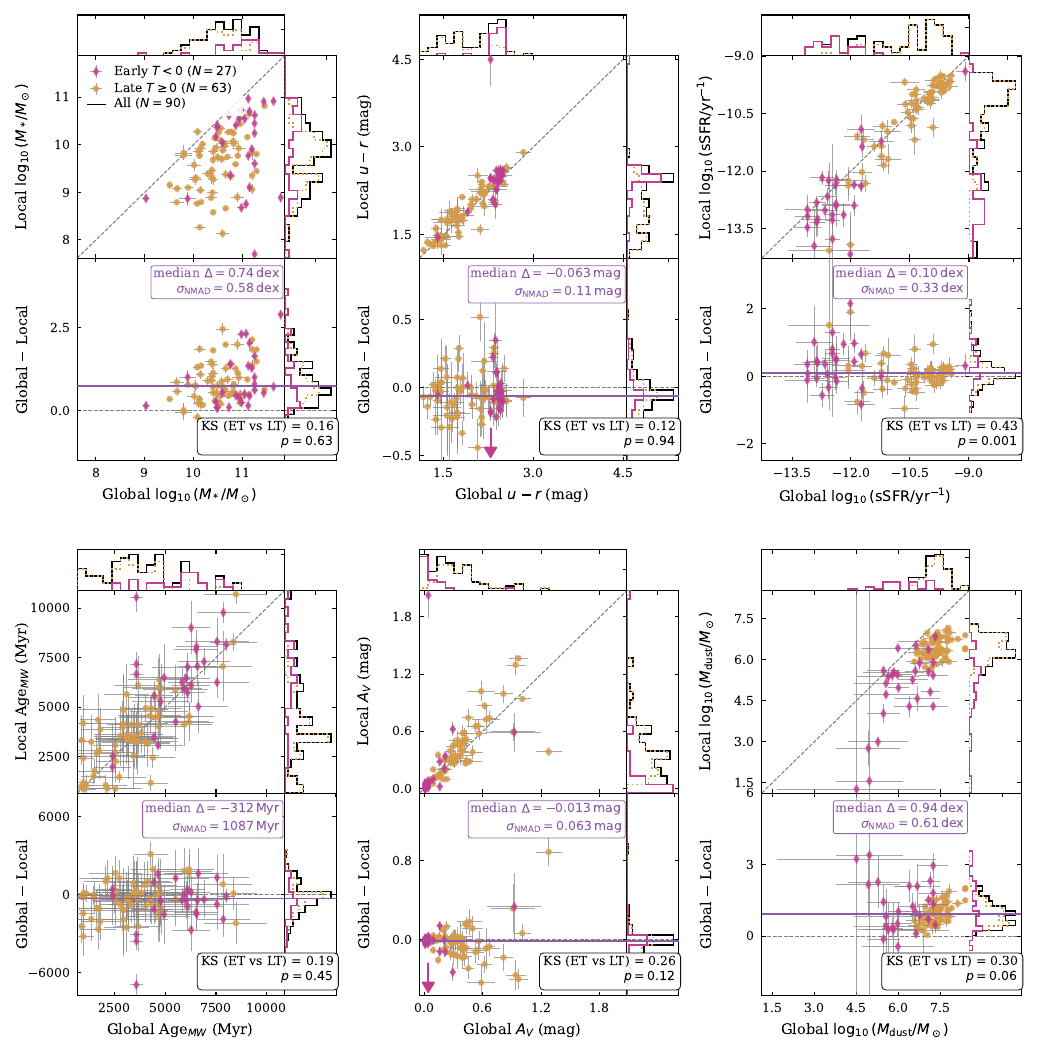}
    \caption{\revision{Local versus global Bayesian estimates from \texttt{CIGALE} SED fitting for 90 SNe Ia across 78 DustPedia host galaxies. In each panel, the upper plot compares the property measured within a 3\,kpc aperture centred on the SN site (local, $y$-axis) against the host galaxy value (global, $x$-axis), with $1\sigma$ uncertainties and the dashed line marking the 1:1 relation; the lower plot shows the global$-$local difference against the global value, with the dashed line marking zero, i.e., equal local and global values. \newrevision{In the $u-r$ and $A_V$ residual panels, SN 1980I (NGC 4374) lies outside the plotted range and is marked by an arrow at its global value ($\Delta$ = $-2.2$ and $-2.0$\,mag respectively). It is included in all quoted statistics.} Points are coded by host galaxy morphology: early-type ($T < 0$, pink, diamonds, solid line histogram) and late-type ($T \geq 0$, gold, circles, dotted histogram), with the full sample as the solid black histogram. Histograms show the distributions of global values (top), local values (upper right) and global$-$local differences (lower right). The solid purple line marks the median of the global$-$local difference for the full sample, annotated together with the scatter $\sigma_{\mathrm{NMAD}}$; a two-sample KS statistic and $p$-value comparing the early- and late-type difference distributions is given in each lower panel. A Wilcoxon signed-rank test finds the median difference inconsistent with zero for every property (\sref{environments}).}}
    \label{fig:lvg_resi}
\end{figure*}

To explore the differences between global and local properties within the SN Ia host galaxies, global vs local values are first investigated directly for each property as displayed in \fref{fig:lvg_resi}. The sample is additionally split at $T = 0$ (using the reported Hubble stage from the DustPedia archive) into early-type \revision{($T < 0$, $N = 27$)} and late-type \revision{($T \geq 0$, $N = 63$)} galaxies to investigate if there are any trends with morphology. $M_*$ and $M_\mathrm{dust}$, as expected, lie systematically below the one-to-one line, with local values lower than global. The other properties, $u-r$, $\mathrm{sSFR}$, $\textrm{Age}_{MW}$ and $A_V$ behave differently, showing correlation but with scatter about the one-to-one line, and no clear offset. The split on morphology is as expected, with the early- and late-type distributions differing at p < 0.01 in every case, for both the local and global measurements. For example, early-type hosts occupy the redder $u-r$ and lower $\textrm{sSFR}$ regions than the late-type hosts. Notably, early-type hosts have higher global $M_*$ than late-type hosts, yet their local $M_*$ span a comparable range, so the clear separation by morphology seen globally is largely washed out within the fixed 3\,kpc aperture. This is largely unsurprising: since SNe Ia occur in galaxies of all types, they also sample the full range of sub-galactic environments available within those galaxies.

\begin{figure*}
    \centering
    \includegraphics[width=\linewidth]{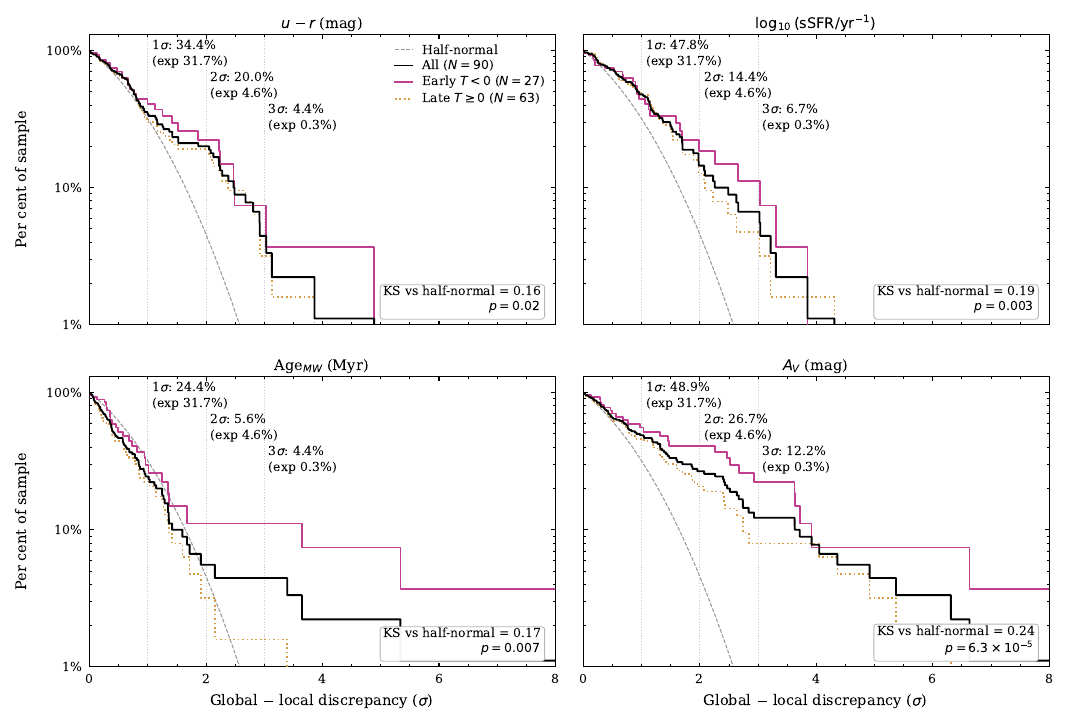}
    \caption{Complementary cumulative distributions of the global-local discrepancy, for $u-r$, $\mathrm{sSFR}$, $\textrm{Age}_{MW}$ and $A_V$. At any point on the $x$-axis, the $y$-axis gives the percentage of SNe Ia in the sample whose global and local estimates disagree by more than that number of combined standard deviations. The dashed grey line shows the half-normal distribution, the expectation if local and global values were identical within measurement uncertainty; if this were the case, 31.7 per cent of the sample would exceed 1$\sigma$, 4.6 per cent would exceed 2$\sigma$, and 0.3 per cent would exceed 3$\sigma$. The sample is split by host galaxy morphology: early-type ($T < 0$, solid pink) and late-type ($T \geq 0$, dotted gold), with the full sample in solid black. Black text annotations indicate what percentages of the full sample exceed 1, 2 and 3$\sigma$ for each property. A KS test comparing the full-sample distribution to the half-normal is given in each panel.}
    \label{fig:pulls}
\end{figure*}

The global-local differences for each property \revision{as a function of the global value are presented in the lower panels of \fref{fig:lvg_resi}}. Unlike in later discrepancy tests (\fref{fig:pulls}), in this investigation I do not take the absolute value of the difference as I wish to see any directionality of the trends. This is clearest, as expected, for both $M_*$ and $M_\mathrm{dust}$ where global exceeds local for the vast majority of the hosts. For the other properties, the directionality is less extreme, but the trends still indicate that global and local properties are providing different information. For $u-r$, SNe Ia have a preference for redder local environments than their host-galaxy average, \revision{with a median offset of $-0.063$\,mag ($\sigma_{\textrm{NMAD}} = 0.11$\,mag),} the opposite of the trend found by \citet{Kelsey2021} for the 3-year DES sample, with the main difference being the redshift range investigated. \revision{The remaining properties show systematic offsets towards lower local $\mathrm{sSFR}$ ($+0.10$\,dex, $\sigma_\mathrm{NMAD} = 0.33$\,dex), higher local $A_V$ ($-0.013$\,mag, $\sigma_\mathrm{NMAD} = 0.063$\,mag) and older local $\textrm{Age}_{MW}$ ($-312$\,Myr, $\sigma_\mathrm{NMAD} = 1087$\,Myr). In each case a Wilcoxon signed-rank test rejects a distribution centred on zero ($p < 0.01$), so these shifts are systematic, although small compared to the object-to-object scatter.} When splitting on morphology, a Kolmogorov–Smirnov (KS) test finds that the early- and late-type difference distributions differ for $\textrm{sSFR}$ \revision{($p=0.001$)}, but are consistent for the other properties \revision{(all $p>0.05$)}. \revision{As the 3\,kpc local aperture is a fixed physical size, its angular radius varies across the sample (see \fref{fig:ap_vs_dist} and \fref{fig:eg_apertures}); to confirm this introduces no effect, the differences were also examined against the redshift-independent host distance, with no property showing a correlation from a Spearman rank test.}

In \fref{fig:pulls} complementary cumulative distributions of the discrepancies between global and local measurements are presented for $u-r$, $\mathrm{sSFR}$, $\textrm{Age}_{MW}$ and $A_V$, and compared to what one might expect if any differences were due to measurement uncertainties alone, to better evaluate the above findings. A $\sigma$ discrepancy is defined as:
\begin{equation}
    \sigma = \frac{|\text{global} - \text{local}|}{\sqrt{\sigma_\text{local}^2 + \sigma_\text{global}^2}}
\end{equation}
This measures how many standard deviations apart the local and global parameters are. If the fitted values from \texttt{CIGALE} have well-constrained uncertainties, and if global and local were consistent within these uncertainties, this discrepancy value should follow a half-normal distribution (the absolute value of a normal distribution). If that were true, one would expect 68.3 per cent of global-local discrepancies to be within $1\sigma$ (meaning 31.7 per cent exceed $1\sigma$), 95.4 per cent $<2\sigma$ and 99.7 per cent $<3\sigma$. Thus, in \fref{fig:pulls}, if the distribution lies above the trend expected for a half-normal distribution, showing an excess of the sample having a discrepancy larger than a given $\sigma$, it indicates that the local SN Ia environments differ from the host galaxy average. For example, if 50 per cent of the sample lies above 1$\sigma$, half of all SN Ia sites have a local value that differs from the host galaxy value by more than the combined measurement uncertainty. For properties such as $M_*$ and $M_\mathrm{dust}$ that depend on area, large discrepancies between global and local are expected (as illustrated in \fref{fig:lvg_resi}), and are thus not as informative as those for $u-r$, $\mathrm{sSFR}$, $\textrm{Age}_{MW}$ and $A_V$ where large discrepancies would indicate clear environmental variation.

As can be seen, for $u-r$, $\mathrm{sSFR}$ and $A_V$, \revision{the distributions lie well above the half-normal expectation, showing} clearly that global and local properties are more different than would be expected from \revision{measurement uncertainties alone. In the $\mathrm{sSFR}$ and $A_V$ panels the early-type hosts show a larger excess than the late-type hosts, although with only 27 early-type SNe I am cautious about over-interpreting this split.} Looking at the $A_V$ panel in more detail, it can be seen that it has the strongest discrepancy of the parameters measured, and a KS test strongly rejects \revision{($p=6.3\times10^{-5}$)} the half-normal. This suggests that local $A_V$ varies more from the global value than the other properties. \revision{I note that the $A_V$ differences are small in absolute terms ($\sigma_\mathrm{NMAD} = 0.063$\,mag; \fref{fig:lvg_resi}), but the \texttt{CIGALE} uncertainties on $A_V$ are correspondingly small, so the differences are large relative to the measurement precision.} If the SN Ia luminosity host-dependence is truly due to dust in the SN host galaxy, the relevant dust will likely be local to the SN (perhaps even circumstellar), and a global host $A_V$ is a poor proxy for it.

Interestingly, the picture for $\textrm{Age}_{MW}$ is quite different. \revision{A KS test formally rejects the half-normal ($p=0.007$), but the trend is in the opposite sense to the other properties at small discrepancies: only 24.4 per cent of the sample exceeds $1\sigma$ against an expected 31.7 per cent, meaning global and local $\textrm{Age}_{MW}$ agree often.} As displayed in \fref{fig:lvg_resi}, $\textrm{Age}_{MW}$ values have some of the largest uncertainties of the properties measured in this analysis, which may be suppressing the discrepancy values presented. \revision{The excess at large discrepancies is driven by a subset of early-type hosts, and given the sample sizes, I am cautious about ascribing meaning to the differences between the early- and late-type distributions.}
\section{SN Ia Siblings}\label{siblings}

In the sample of SNe Ia within DustPedia galaxies that pass selection cuts, there are 10 galaxies hosting multiple SNe Ia, referred to as ``siblings'' \citep{Brown2014}. By being associated with the same host galaxy, sets of siblings share global properties, alongside uncertainties from redshift, peculiar velocities, and gravitational lensing. These sibling environments provide excellent case-studies to better illuminate the key difference between local and global host galaxy parameters for SNe Ia, and could determine if sub-galactic differences are the dominant environmental factor requiring standardisation in SN Ia cosmology \citep{Kelsey2024}. 

\begin{figure*}
    \centering
    \includegraphics[width=\linewidth]{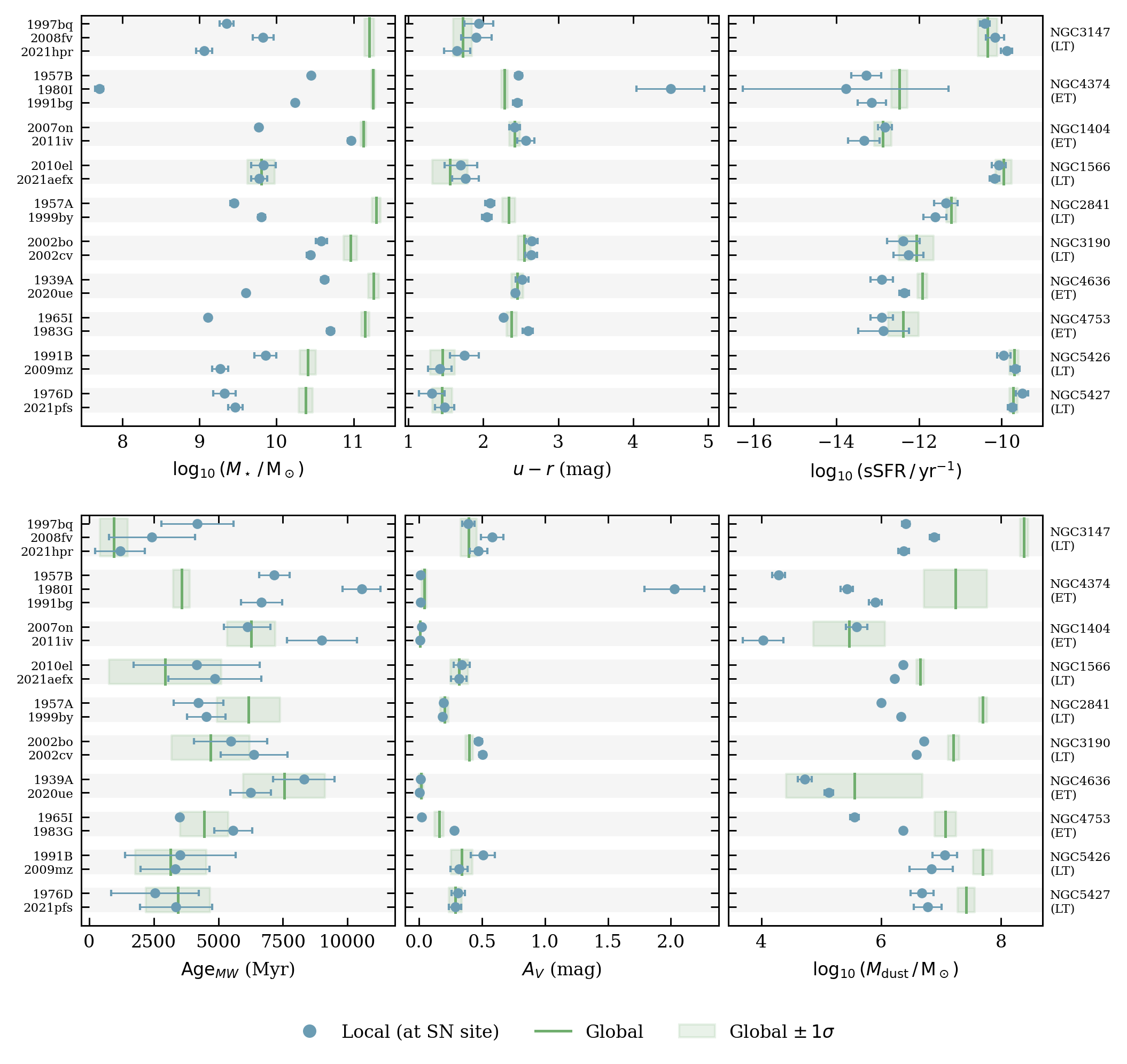}
    \caption{A subset of local (3\,kpc aperture) properties at each SN Ia site compared to the global host galaxy value for the 10 sibling hosts in the sample. Each row corresponds to one host galaxy, with individual SNe labelled on the left axis and host names on the right. The solid green line marks the global Bayesian estimate from \texttt{CIGALE}, with the shaded band showing the $\pm\,1\sigma$ uncertainty. Blue points are the local measurements at each SN site with associated uncertainties. Whether the galaxy is Early-Type (ET) or Late-Type (LT) is indicated in brackets under the name.}
    \label{fig:sibs_lvg}
\end{figure*}

In \fref{fig:sibs_lvg}, the differences in local and global properties from \texttt{CIGALE} are presented for the sibling sample, \revision{with the values presented in \tref{tab:siblings_local_global}}. The properties that are dependent on area are as one would expect, e.g. local $M_*$ are all below the global $M_*$ for the sample (with the exception of NGC\,1566, where the local regions span a large extent of this galaxy, with significant spatial overlap between the siblings), and likewise for $M_\mathrm{dust}$. \revision{Of the 22 sibling SNe, 20 (91 per cent) have local and global $M_*$ whose error bars do not overlap, that is, they differ by more than the sum of their uncertainties; the equivalent for $M_\mathrm{dust}$ is 19/22 (86 per cent).} However, a clear spread in values can be seen between siblings within the same host, reflecting the spatial variation across the galaxy. To put this into context for SN Ia cosmology, despite almost all being associated with high-mass galaxies and thus being assigned to the same side of the global mass step bin (split at $\log(M_*/\mathrm{M_\odot}) = 10$) for standardisation, there are a few cases where these sets of siblings would fall on opposite sides of a local mass step split point at $\log(M_*/\mathrm{M_\odot})=9$ \citep[as defined in e.g.,][]{Kelsey2021,Kelsey2023}. 

\begin{figure*}
    \centering
    \includegraphics[width=\linewidth]{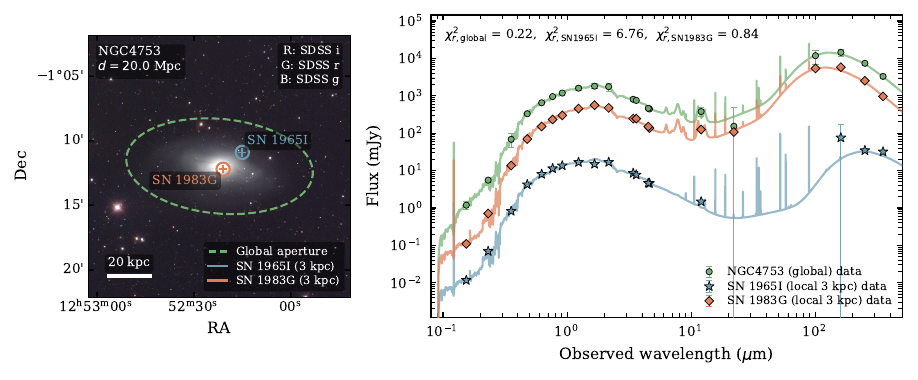}\\[2pt]
    \includegraphics[width=\linewidth]{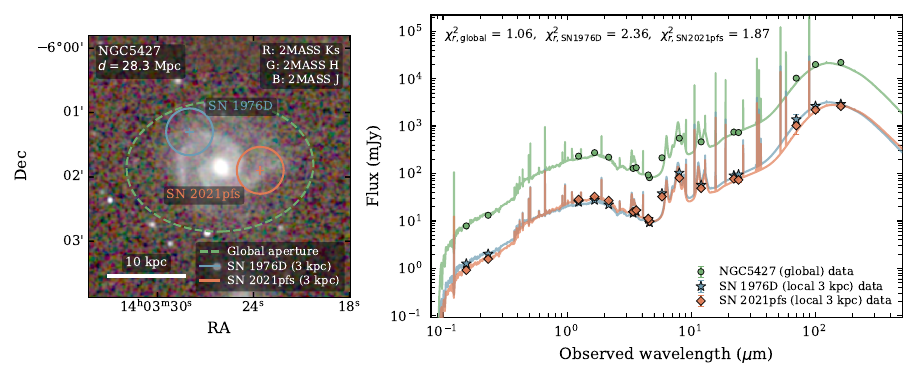}
    \caption{\textit{Top:} Two-panel view of the sibling SN Ia host NGC\,4753. The left panel shows a three-colour SDSS $gri$ composite, with the dashed green ellipse marking the DustPedia global photometric aperture and solid circles showing the 3\,kpc local apertures at the sites of SN\,1965I (blue) and SN\,1983G (orange). The right panel shows the best-fit \texttt{CIGALE} SEDs for the global galaxy (green) and each local aperture, with observed photometry overplotted and reduced $\chi^2$ values indicated. \textit{Bottom:} As above but for NGC\,5427. The three-colour composite uses 2MASS $JHK_s$ due to the lack of available SDSS data for this galaxy.}
    \label{fig:sibling_panels}
\end{figure*}

For the other properties ($u-r$, $\mathrm{sSFR}$, $\textrm{Age}_{MW}$, $A_V$) local measurements that fall outside the global error band indicate that the galaxy-averaged value does not capture the true conditions at each SN site. \revision{Applying the same criterion to these properties gives 7/22 (32 per cent) in $u-r$, 5/22 (23 per cent) in $\mathrm{sSFR}$, 5/22 (23 per cent) in $\textrm{Age}_{MW}$ and 9/22 (41 per cent) in $A_V$. These fractions are lower limits as the local and global apertures are not independent; the 3\,kpc region lies within the global, so summing their uncertainties overstates the error on the difference.} The siblings within NGC\,4753 (SN\,1965I and SN\,1983G), a lenticular galaxy with distinct dust lanes, have a striking difference in many properties, despite having comparable $\mathrm{sSFR}$ \revision{with error bars overlapping} the uncertainties for the galaxy value. Notably, they have very different dust conditions e.g. attenuation ($A_V = 0.018\pm0.003$ vs $0.280\pm0.018$\,mag), dust luminosities ($\textrm{log}(L_{\textrm{dust}}/{L_\odot}) = 6.88\pm0.06$ vs $9.45\pm0.02$), radiation field intensities ($U_{\textrm{min}} = 0.11\pm0.02$ vs $5.96\pm0.58$), and hydrocarbon fractions ($q_\mathrm{hac} = 0.037\pm0.027$ vs $0.067\pm0.015$)\footnote{\revision{See \tref{tab:siblings_local_global} for the full set of sibling environment parameters.}}. Similarly, their mass-weighted ages sit at the furthest extents of the uncertainties for the galaxy value. Considering these two SNe to come from comparable environments is clearly incorrect. Looking at the galaxy itself, the reason becomes quite obvious, as displayed in \fref{fig:sibling_panels}. SN\,1983G is close to the centre of the galaxy, with much of its local aperture including the light from the galaxy centre. SN\,1965I, on the other hand, is offset towards the outskirts of the galaxy. This means that the SED for \revision{the local environment of} SN\,1983G is much more similar to the galaxy average than \revision{the local environment of} SN\,1965I \revision{is}.

Conversely, there are some situations where, despite the siblings being on opposite sides of the galaxy, their environmental properties are similar. SN\,1976D and SN\,2021pfs are both located at the edges of the spiral arms of NGC\,5427, roughly the same distance from the core, and have near-identical SEDs as shown in \fref{fig:sibling_panels}. Across the board, the properties of their local environments are within the uncertainties of one another, and are consistent with their galaxy average values. 

Some of these siblings have been previously investigated in the literature. \citet{Elias-Rosa2008} found evidence of SN\,2002cv in NGC\,3190 being more heavily dust obscured than its sibling SN\,2002bo. From the SN light-curves, they report an $A_V$ value of $\sim 8$ for SN\,2002cv, with corresponding $R_V = 1.59$, notably different to the $R_V = 3.1$ value for SN\,2002bo. The $A_V$ reported by \citet{Elias-Rosa2008} is a line of sight host extinction value to a point source whilst the local $A_V$ values reported in this analysis are integrated across the 3\,kpc local aperture region for each SN. These quantities are fundamentally different and should not be used as proxies for one another, despite the nomenclature; see \citet{Duarte2025} for a detailed analysis alongside discussion in \sref{Discussion}. The $A_V$ values from this analysis are comparable for each of the NGC\,3190 siblings \revision{(SN\,2002bo $A_V = 0.471 \pm 0.027$ and SN\,2002cv $A_V = 0.503 \pm 0.026$), both} considerably lower than the line of sight value reported for SN\,2002cv. These differences may indicate the presence of circumstellar dust around SN\,2002cv which cannot be measured from even this local environmental analysis. 
\section{Discussion}\label{Discussion}

The case for measuring SN Ia environments locally rests on the assumption that the local environment differs from the host average, and that the difference matters for the SN. This analysis tests the first part of that assumption directly. For $u-r$, $\textrm{sSFR}$, and $A_V$, the value within 3\,kpc of the SN differs from the host galaxy average by more than the combined uncertainties, with $A_V$ showing the largest discrepancy \revision{(rejecting the half-normal expectation at $p=6.3\times10^{-5}$)}. The sibling hosts tie this offset clearly to local differences, as any difference between their environments can only come from the part of the galaxy each one exploded in. As an example, the two SNe in NGC\,4753 fall on opposite sides of the host value for $u-r$, $A_V$ and $\textrm{Age}_{MW}$, beyond the extent of the uncertainties in these global measurements. The divergence is not universal across the sample, since siblings sampling similar structure agree (NGC\,5427); however where this divergence occurs, a single host-averaged value cannot describe multiple SN sites at once. There is local information a host-averaged measurement cannot capture. 

The environmental property estimates benefit from the broad wavelength coverage of the DustPedia data, extending into the FIR, which helps to \revision{partially break} the age-dust degeneracy that limits the optical- and UV-based SED fits used in much of the hosts literature \citep[see recent discussion in][]{Kim2024,Ramaiya2025,Murakami2026}. Even so, recovering stellar ages from broadband photometric SED fitting remains difficult, given the known sensitivities to the choice of star-formation history \citep{Carnall2019, Leja2019}. A clearer statement about local versus global ages for SN Ia hosts will therefore require a larger sample and age estimates less reliant on photometry alone, such as those from integral-field-unit (IFU) spectroscopy \citep[e.g.,][]{Galbany2018} \revision{or dedicated high signal-to-noise spectroscopy \citep[e.g.,][]{Kang2016,Kang2020}}.

When the environmental relationship is attributed to ``dust'', the word is describing two distinct physical quantities. The extinction measured from a SN light curve refers to a single line of sight, whereas the dust inferred for a galaxy, or for an aperture within one, is an attenuation: an integral over many lines of sight that folds in the relative geometry of stars and dust together with the effects of scattering \citep{Chevallard2013, Narayanan2018, Duarte2023}. The hosts literature too often reports the two interchangeably, when the distinction needs to remain clear. The $A_V$ reported in this analysis is of the latter kind, the attenuation of the stellar population within the aperture, not the extinction the SN light actually traverses, nor a reliable proxy for it \citep{Popovic2024,Duarte2025}. My results sharpen the problem: global $A_V$ is the poorest predictor of its local counterpart of any property measured, even setting the extinction-attenuation mismatch aside, a host-averaged dust measurement already misses local information. 

\section{Summary and Conclusions}\label{Summary}

Using DustPedia imaging and \texttt{CIGALE} SED fitting, I compared the local environments of a sample of \revision{90} nearby SNe Ia, measured within a fixed 3\,kpc aperture at each SN site, against the global properties of their host galaxies, to test whether the sub-galactic environment carries different information to the global average, relevant to ongoing discussions over the optimal SN Ia environmental standardisation. By using imaging spanning from the UV to the FIR, I am able to better constrain the dust properties of each environment and partially break the age-dust degeneracy that limits studies using only optical-based photometry. Such coverage is seldom used in SN Ia host galaxy studies, with the work of \citet{Ramaiya2025,Ramaiya2026} being recent exceptions on global scales, and further work forthcoming (Tweddle et al., in prep.), but to my knowledge it has not previously been applied to sub-galactic regions around SNe Ia.  

The key findings are as follows:
\begin{itemize}
    \item For $u-r$, $\textrm{sSFR}$ and $A_V$, local and global values differ by more than their combined measurement uncertainty, with $A_V$ showing the strongest discrepancy \revision{relative to its measurement uncertainties (rejecting the half-normal expectation at $p=6.3\times10^{-5}$). Each also shows a small systematic offset, towards redder, more quiescent and more attenuated local environments than the host average.} This reflects spatial variation across a host galaxy, unlike $M_*$ and $M_\mathrm{dust}$, which are lower locally simply because the local aperture encloses less of the galaxy.
    \item \revision{The local$-$global differences in $\textrm{sSFR}$ depend on morphology ($p=0.001$), with early-type SN sites tending to be more quiescent than their host average; the early-type hosts also show a larger excess in the $A_V$ discrepancy distribution. With only 27 SNe in early-type hosts, this may reflect small-number statistics.}
    \item \revision{Local and global $\textrm{Age}_{MW}$ are consistent within their measurement uncertainties for the majority of the sample, with a small systematic offset towards older local ages ($-312$\,Myr, well within the typical per-object uncertainty).}
    \item In the sibling hosts, where the global properties are fixed within a set, the variation in properties between sibling sites depends on the set in question. Siblings sampling similar structure in the galaxy are consistent, whilst others differ considerably.
\end{itemize}

This approach involves a deliberate compromise. The 3\,kpc aperture matches the local scale adopted by many higher-redshift cosmological surveys, but recovering it alongside the FIR coverage necessary is only possible for galaxies as nearby and well-resolved as those in DustPedia. Sharper spatial resolution could be obtained by dropping the longer-wavelength bands with poorest angular resolution, but at the cost of the FIR information central to several of the properties measured. The absence of comparable resolved UV-to-FIR data at higher redshift is the barrier to a similar locally-resolved analysis across a cosmological sample, which would require high-resolution IR data. Whilst \textit{JWST} will help in part, since its NIR and MIR resolution would let us resolve local environments to higher redshift, it does not reach the FIR. 

The sibling environments show that there is no single prescription for scaling a global property to a local counterpart. \revision{This is seen across the sample: the median global$-$local offsets are a fraction of their object-to-object scatter (e.g. $-0.063$ vs $\sigma_\mathrm{NMAD} = 0.11$\,mag in $u-r$; $-0.013$ vs $0.063$\,mag in $A_V$), so any single correction would be smaller than the spread it aims to capture.} The offset \revision{likely} depends on morphology, inclination, and on where each SN sits within its galaxy, before even reaching subtler differences in star formation history, dust composition or merger history. A proxy capturing all of this would need so many parameters per individual SN that measuring the local environment directly is the simpler option.

Many local environment properties clearly depart from the global host values, meaning any galaxy-integrated property does not capture the conditions at the SN site. By applying global-based corrections, we may be unintentionally masking the true drivers of the SN Ia brightness-environment relation. How these environmental properties compare with light-curve standardisation properties across this sample will be the focus of the second paper in this series.

\section*{Acknowledgements}

\revision{I thank the anonymous referee for their careful reading and constructive comments, which have improved this paper.}

I acknowledge support for an Early Career Fellowship from the Leverhulme Trust through grant ECF-2024-054 and the Isaac Newton Trust through grant 24.08(w). I thank the DustPedia team for making their data publicly available, and am grateful to Christopher Clark for generously providing the DustPedia imaging directly while the archive was temporarily offline.

This work was performed using resources provided by the Cambridge Service for Data Driven Discovery (CSD3) operated by the University of Cambridge Research Computing Service (www.csd3.cam.ac.uk), provided by Dell EMC and Intel using Tier-2 funding from the Engineering and Physical Sciences Research Council (capital grant EP/T022159/1), and DiRAC funding from the Science and Technology Facilities Council (www.dirac.ac.uk).

DustPedia is a collaborative focused research project supported by the European Union under the Seventh Framework Programme (2007-2013) call (proposal no. 606847). The participating institutions are: Cardiff University, UK; National Observatory of Athens, Greece; Ghent University, Belgium; Université Paris Sud, France; National Institute for Astrophysics, Italy and CEA, France.

This research made use of Photutils, an Astropy package for
detection and photometry of astronomical sources \citep{Bradley2025}.

Anthropic’s Claude Opus 4.8 has been used to aid proofreading, fix code bugs, and suggest code optimisations and plot improvements. 

\section*{Data Availability}

The imaging data and host photometry associated with this paper are all publicly available through the DustPedia project at \url{http://dustpedia.astro.noa.gr/} \revision{and \url{https://doi.org/10.26131/IRSA660}.}

\revision{Local aperture photometry for the 3\,kpc fiducial apertures and the 1\,kpc appendix test are presented in the online supplementary material, alongside \texttt{CIGALE} best-fit parameters and uncertainties for local and global environments. The values for the sibling sample are additionally presented in \tref{tab:siblings_local_global}.}



\bibliographystyle{mnras}
\bibliography{biblio} 



\appendix
\section{Testing a 1\,kpc radius aperture} \label{aptest}
\begin{figure*}
    \centering
    \includegraphics[width=\linewidth]{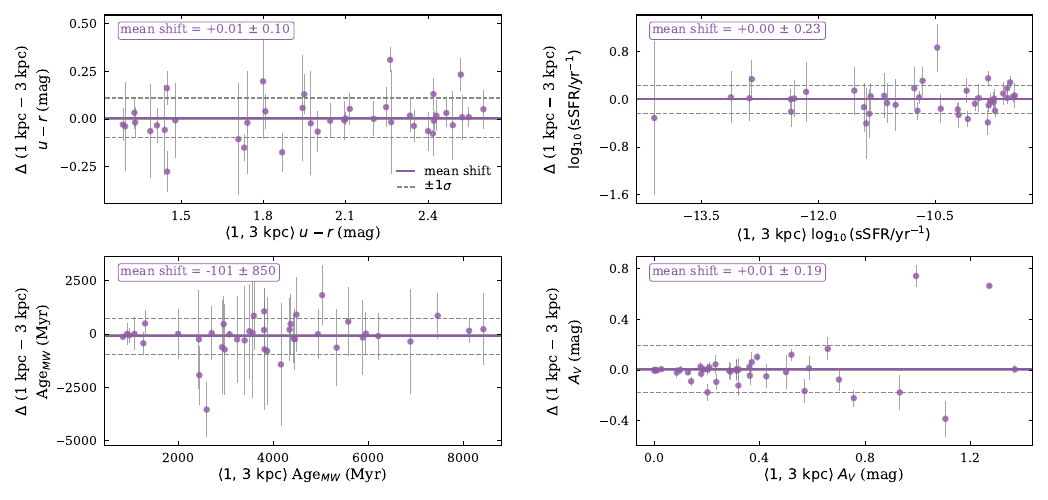}
    \caption{Difference between the local \texttt{CIGALE} estimates at 1\,kpc and 3\,kpc (1\,kpc minus 3\,kpc), plotted against the mean of the two apertures, for the $41$ SNe fitted at both radii. In each panel the solid purple line marks the mean shift between the apertures with the dashed lines indicating $\pm 1\sigma$.}
    \label{fig:ap_means}
\end{figure*}

The local apertures used throughout this analysis have a radius of 3\,kpc. Whilst a smaller radius samples the SN environment more closely, it may be below the $\sigma_\textrm{PSF}$ of the data for more distant hosts, making the photometry unreliable. To test whether the comparison between local and global properties depends on the adopted local radius, I repeated the local \texttt{CIGALE} fits at 1\,kpc and compared the two sets of recovered environmental parameters for the \revision{41} SNe that pass the sample cuts of \sref{methods} and whose 1\,kpc aperture remains above $\sigma_\textrm{PSF}$ for the convolved data.

As expected, the stellar mass and dust mass are both lower at 1\,kpc than 3\,kpc, by a mean of \revision{$0.90 \pm 0.24$} and \revision{$1.12 \pm 1.12$\,dex} respectively, resulting from the smaller aperture enclosing less light. The $u-r$, $\mathrm{sSFR}$, $\textrm{Age}_{MW}$, and $A_V$ are unchanged within the uncertainties and have mean differences consistent with zero, as shown in \fref{fig:ap_means}; \revision{none differs from zero under a Wilcoxon signed-rank test}. I note that for $A_V$, the differences appear largest for those with the highest mean $A_V$ values. 

\begin{figure*}
    \centering
    \includegraphics[width=\linewidth]{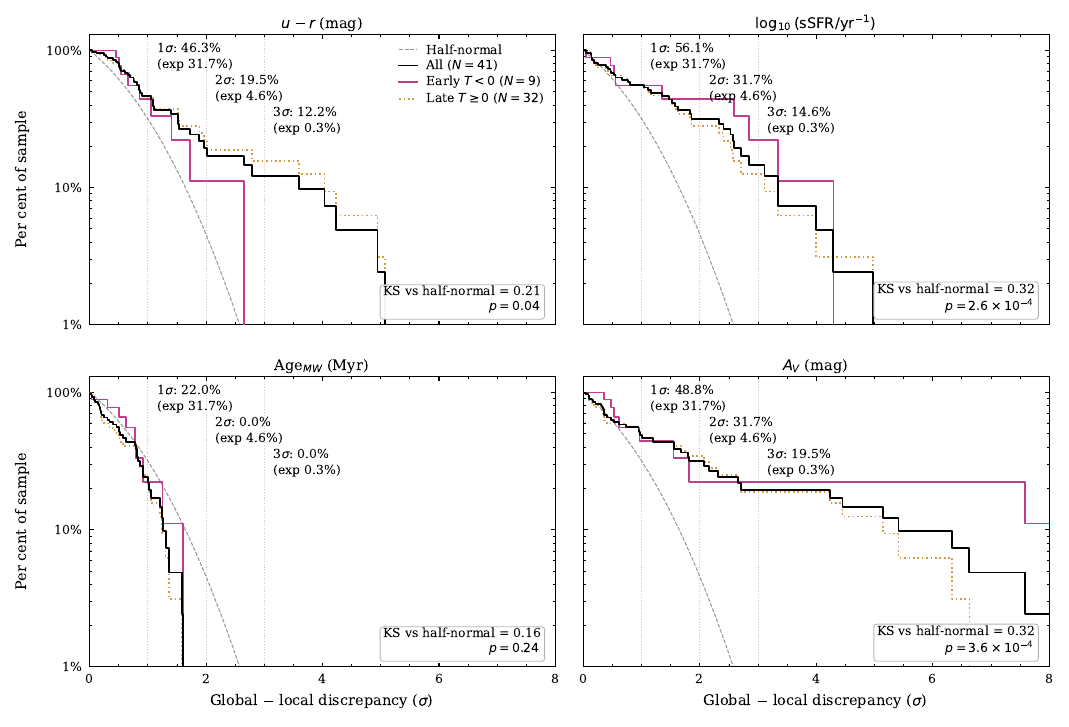}
    \caption{As \fref{fig:pulls}, but for a 1\,kpc local region around each SN.}
    \label{fig:pulls-1kpc}
\end{figure*}

The local-to-global differences reported in \sref{environments} are reproduced at 1\,kpc. Repeating the analysis of the discrepancy between global and local measurements with the 1\,kpc local values as shown in \fref{fig:pulls-1kpc}, the results are largely consistent with what was found for 3\,kpc. The $\mathrm{sSFR}$ and $A_V$ distributions depart from the half-normal expectation at \revision{$p = 2.6\times10^{-4}$ and $p = 3.6\times10^{-4}$}, indicating that global and local properties are more different than would be expected from measurement uncertainties alone. \revision{The departure for $u-r$ is larger than at 3\,kpc (46 against 34 per cent beyond $1\sigma$; KS $D = 0.21$ against $0.16$) but $p = 0.04$ with 41 objects.} The $\textrm{Age}_{MW}$ is consistent with global and local values being within the uncertainties for each other (\revision{i.e., consistent with the half-normal expectation, $p = 0.24$}), with no SN showing more than $2\sigma$ discrepancy between local and global values.

\section{Sibling Properties}

\tref{tab:siblings_local_global} lists the local (3\,kpc aperture) and global \texttt{CIGALE} estimates for the 22 SNe\,Ia in the 10 sibling host galaxies, as presented in \fref{fig:sibs_lvg}. Equivalent values for the full sample of \revision{90} SNe\,Ia are available as online supplementary material.

\onecolumn
\begin{landscape}
\setlength{\LTcapwidth}{\textwidth}
\begin{longtable}{llccccccccc}
\caption{Local (3\,kpc) and global \texttt{CIGALE} properties of the SN\,Ia sibling environments.}\label{tab:siblings_local_global} \\
\hline
Galaxy & Aperture & $\log(M_*/\mathrm{M_\odot})$ & $u - r$ & $\log(\mathrm{sSFR}/\mathrm{yr}^{-1})$ & Age & $A_V$ & $\log(M_\mathrm{dust}/\mathrm{M_\odot})$ & $\log(L_\mathrm{dust}/\mathrm{L_\odot})$ & $U_\mathrm{min}$ & $q_\mathrm{HAC}$ \\
 & & & (mag) & & (Myr) & (mag) & & & & \\
\hline
\endfirsthead
\hline
Galaxy & Aperture & $\log(M_*/\mathrm{M_\odot})$ & $u - r$ & $\log(\mathrm{sSFR}/\mathrm{yr}^{-1})$ & Age & $A_V$ & $\log(M_\mathrm{dust}/\mathrm{M_\odot})$ & $\log(L_\mathrm{dust}/\mathrm{L_\odot})$ & $U_\mathrm{min}$ & $q_\mathrm{HAC}$ \\
 & & & (mag) & & (Myr) & (mag) & & & & \\
\hline
\endhead
\hline
\endfoot
NGC\,3147 & Global & $11.20 \pm 0.06$ & $1.72 \pm 0.13$ & $-10.34 \pm 0.23$ & $961 \pm 540$ & $0.394 \pm 0.062$ & $8.39 \pm 0.07$ & $11.05 \pm 0.03$ & $2.12 \pm 0.51$ & $0.128 \pm 0.023$ \\
 & SN 1997bq & $9.35 \pm 0.09$ & $1.94 \pm 0.19$ & $-10.40 \pm 0.12$ & $4190 \pm 1401$ & $0.391 \pm 0.050$ & $6.41 \pm 0.06$ & $8.84 \pm 0.03$ & $1.23 \pm 0.27$ & $0.138 \pm 0.015$ \\
 & SN 2008fv & $9.83 \pm 0.14$ & $1.91 \pm 0.20$ & $-10.16 \pm 0.22$ & $2432 \pm 1668$ & $0.579 \pm 0.090$ & $6.89 \pm 0.08$ & $9.68 \pm 0.04$ & $2.97 \pm 0.76$ & $0.120 \pm 0.020$ \\
 & SN 2021hpr & $9.06 \pm 0.10$ & $1.65 \pm 0.17$ & $-9.88 \pm 0.13$ & $1199 \pm 960$ & $0.471 \pm 0.071$ & $6.38 \pm 0.09$ & $9.11 \pm 0.02$ & $2.57 \pm 0.73$ & $0.135 \pm 0.015$ \\
\noalign{\vskip 2pt}
NGC\,4374 & Global & $11.26 \pm 0.02$ & $2.28 \pm 0.04$ & $-12.47 \pm 0.19$ & $3576 \pm 319$ & $0.040 \pm 0.018$ & $7.24 \pm 0.53$ & $9.37 \pm 0.20$ & $6.06 \pm 15.72$ & $0.293 \pm 0.098$ \\
 & SN 1957B & $10.45 \pm 0.02$ & $2.47 \pm 0.05$ & $-13.28 \pm 0.36$ & $7168 \pm 593$ & $0.013 \pm 0.002$ & $4.29 \pm 0.11$ & $7.83 \pm 0.06$ & $9.05 \pm 2.52$ & $0.250 \pm 0.085$ \\
 & SN 1980I & $7.70 \pm 0.05$ & $4.50 \pm 0.45$ & $-13.77 \pm 2.49$ & $10543 \pm 735$ & $2.028 \pm 0.238$ & $5.43 \pm 0.10$ & $6.84 \pm 0.04$ & $0.15 \pm 0.06$ & $0.041 \pm 0.021$ \\
 & SN 1991bg & $10.24 \pm 0.03$ & $2.45 \pm 0.06$ & $-13.14 \pm 0.35$ & $6674 \pm 787$ & $0.013 \pm 0.002$ & $5.91 \pm 0.11$ & $7.63 \pm 0.06$ & $0.13 \pm 0.06$ & $0.260 \pm 0.088$ \\
\noalign{\vskip 2pt}
NGC\,1404 & Global & $11.13 \pm 0.04$ & $2.42 \pm 0.07$ & $-12.87 \pm 0.20$ & $6273 \pm 922$ & $0.009 \pm 0.002$ & $5.47 \pm 0.60$ & $8.37 \pm 0.10$ & $11.26 \pm 20.01$ & $0.136 \pm 0.086$ \\
 & SN 2007on & $9.77 \pm 0.03$ & $2.42 \pm 0.07$ & $-12.82 \pm 0.17$ & $6117 \pm 895$ & $0.018 \pm 0.003$ & $5.59 \pm 0.18$ & $7.35 \pm 0.07$ & $0.27 \pm 0.25$ & $0.114 \pm 0.056$ \\
 & SN 2011iv & $10.97 \pm 0.04$ & $2.57 \pm 0.11$ & $-13.33 \pm 0.38$ & $9008 \pm 1352$ & $0.007 \pm 0.002$ & $4.03 \pm 0.34$ & $8.01 \pm 0.09$ & $16.14 \pm 18.89$ & $0.256 \pm 0.103$ \\
\noalign{\vskip 2pt}
NGC\,1566 & Global & $9.80 \pm 0.18$ & $1.55 \pm 0.24$ & $-9.95 \pm 0.19$ & $2945 \pm 2177$ & $0.318 \pm 0.070$ & $6.65 \pm 0.06$ & $9.58 \pm 0.02$ & $3.90 \pm 0.93$ & $0.105 \pm 0.013$ \\
 & SN 2010el & $9.83 \pm 0.16$ & $1.70 \pm 0.22$ & $-10.07 \pm 0.17$ & $4167 \pm 2442$ & $0.339 \pm 0.062$ & $6.37 \pm 0.02$ & $9.51 \pm 0.02$ & $6.00 \pm 0.07$ & $0.101 \pm 0.005$ \\
 & SN 2021aefx & $9.78 \pm 0.10$ & $1.76 \pm 0.18$ & $-10.17 \pm 0.12$ & $4867 \pm 1796$ & $0.315 \pm 0.062$ & $6.22 \pm 0.02$ & $9.37 \pm 0.02$ & $6.00 \pm 0.09$ & $0.103 \pm 0.011$ \\
\noalign{\vskip 2pt}
NGC\,2841 & Global & $11.30 \pm 0.05$ & $2.34 \pm 0.09$ & $-11.22 \pm 0.12$ & $6167 \pm 1217$ & $0.201 \pm 0.032$ & $7.70 \pm 0.07$ & $10.10 \pm 0.03$ & $1.21 \pm 0.25$ & $0.090 \pm 0.037$ \\
 & SN 1957A & $9.45 \pm 0.05$ & $2.09 \pm 0.06$ & $-11.35 \pm 0.29$ & $4233 \pm 964$ & $0.192 \pm 0.021$ & $6.01 \pm 0.04$ & $8.41 \pm 0.02$ & $1.22 \pm 0.16$ & $0.061 \pm 0.008$ \\
 & SN 1999by & $9.80 \pm 0.05$ & $2.05 \pm 0.07$ & $-11.61 \pm 0.28$ & $4536 \pm 748$ & $0.184 \pm 0.019$ & $6.33 \pm 0.03$ & $8.74 \pm 0.02$ & $1.20 \pm 0.06$ & $0.101 \pm 0.008$ \\
\noalign{\vskip 2pt}
NGC\,3190 & Global & $10.96 \pm 0.09$ & $2.55 \pm 0.09$ & $-12.07 \pm 0.42$ & $4702 \pm 1516$ & $0.396 \pm 0.029$ & $7.21 \pm 0.09$ & $9.97 \pm 0.03$ & $2.96 \pm 0.74$ & $0.071 \pm 0.018$ \\
 & SN 2002bo & $10.58 \pm 0.08$ & $2.65 \pm 0.08$ & $-12.38 \pm 0.40$ & $5476 \pm 1415$ & $0.471 \pm 0.027$ & $6.72 \pm 0.04$ & $9.58 \pm 0.02$ & $3.60 \pm 0.49$ & $0.070 \pm 0.017$ \\
 & SN 2002cv & $10.44 \pm 0.04$ & $2.64 \pm 0.08$ & $-12.26 \pm 0.36$ & $6384 \pm 1299$ & $0.503 \pm 0.026$ & $6.59 \pm 0.02$ & $9.45 \pm 0.02$ & $3.49 \pm 0.09$ & $0.099 \pm 0.006$ \\
\noalign{\vskip 2pt}
NGC\,4636 & Global & $11.26 \pm 0.07$ & $2.46 \pm 0.08$ & $-11.92 \pm 0.11$ & $7550 \pm 1584$ & $0.016 \pm 0.017$ & $5.56 \pm 1.14$ & $8.79 \pm 0.45$ & $22.68 \pm 26.76$ & $0.230 \pm 0.116$ \\
 & SN 1939A & $10.62 \pm 0.05$ & $2.52 \pm 0.09$ & $-12.90 \pm 0.27$ & $8310 \pm 1186$ & $0.011 \pm 0.002$ & $4.73 \pm 0.12$ & $7.89 \pm 0.07$ & $4.87 \pm 2.09$ & $0.109 \pm 0.071$ \\
 & SN 2020ue & $9.60 \pm 0.03$ & $2.43 \pm 0.03$ & $-12.36 \pm 0.12$ & $6252 \pm 793$ & $0.003 \pm 0.000$ & $5.13 \pm 0.07$ & $6.46 \pm 0.03$ & $0.11 \pm 0.03$ & $0.048 \pm 0.039$ \\
\noalign{\vskip 2pt}
NGC\,4753 & Global & $11.15 \pm 0.05$ & $2.38 \pm 0.07$ & $-12.38 \pm 0.37$ & $4445 \pm 933$ & $0.158 \pm 0.038$ & $7.07 \pm 0.18$ & $9.79 \pm 0.10$ & $2.93 \pm 1.80$ & $0.095 \pm 0.057$ \\
 & SN 1965I & $9.11 \pm 0.02$ & $2.27 \pm 0.02$ & $-12.91 \pm 0.27$ & $3500 \pm 26$ & $0.018 \pm 0.003$ & $5.56 \pm 0.07$ & $6.88 \pm 0.06$ & $0.11 \pm 0.02$ & $0.037 \pm 0.026$ \\
 & SN 1983G & $10.70 \pm 0.04$ & $2.59 \pm 0.07$ & $-12.86 \pm 0.61$ & $5578 \pm 727$ & $0.280 \pm 0.018$ & $6.37 \pm 0.04$ & $9.45 \pm 0.02$ & $5.96 \pm 0.58$ & $0.067 \pm 0.015$ \\
\noalign{\vskip 2pt}
NGC\,5426 & Global & $10.41 \pm 0.10$ & $1.46 \pm 0.17$ & $-9.70 \pm 0.11$ & $3158 \pm 1378$ & $0.338 \pm 0.086$ & $7.69 \pm 0.16$ & $10.42 \pm 0.02$ & $2.71 \pm 0.93$ & $0.103 \pm 0.011$ \\
 & SN 1991B & $9.86 \pm 0.14$ & $1.75 \pm 0.19$ & $-9.95 \pm 0.16$ & $3529 \pm 2140$ & $0.508 \pm 0.095$ & $7.06 \pm 0.20$ & $9.74 \pm 0.02$ & $2.65 \pm 1.23$ & $0.115 \pm 0.019$ \\
 & SN 2009mz & $9.27 \pm 0.10$ & $1.42 \pm 0.16$ & $-9.67 \pm 0.11$ & $3332 \pm 1335$ & $0.318 \pm 0.067$ & $6.84 \pm 0.36$ & $9.29 \pm 0.03$ & $1.82 \pm 1.11$ & $0.115 \pm 0.019$ \\
\noalign{\vskip 2pt}
NGC\,5427 & Global & $10.38 \pm 0.09$ & $1.45 \pm 0.13$ & $-9.72 \pm 0.09$ & $3439 \pm 1233$ & $0.288 \pm 0.051$ & $7.42 \pm 0.14$ & $10.36 \pm 0.02$ & $4.24 \pm 1.43$ & $0.100 \pm 0.004$ \\
 & SN 1976D & $9.32 \pm 0.14$ & $1.31 \pm 0.17$ & $-9.51 \pm 0.15$ & $2549 \pm 1696$ & $0.309 \pm 0.053$ & $6.68 \pm 0.19$ & $9.50 \pm 0.02$ & $3.51 \pm 1.48$ & $0.100 \pm 0.003$ \\
 & SN 2021pfs & $9.47 \pm 0.10$ & $1.48 \pm 0.13$ & $-9.75 \pm 0.10$ & $3358 \pm 1395$ & $0.286 \pm 0.050$ & $6.78 \pm 0.23$ & $9.42 \pm 0.02$ & $2.47 \pm 1.20$ & $0.101 \pm 0.007$ \\
\end{longtable}
\end{landscape}


\bsp	
\label{lastpage}
\end{document}